\documentclass[12pt]{article}
\usepackage{carmaprop-just}
\usepackage{graphicx}
\newcommand{\my}{\mbox{$M_{\odot}$~yr$^{-1}$}}
\newcommand{\ls}{\mbox{$L_{\odot}$}}
\newcommand{\ms}{\mbox{$M_{\odot}$}}
\newcommand{\gsim}{\raisebox{-.4ex}{$\stackrel{>}{\scriptstyle \sim}$}}
\newcommand{\lsim}{\raisebox{-.4ex}{$\stackrel{<}{\scriptstyle \sim}$}}

\def\arcdeg{\hbox{$^\circ$}}
\def\12co{$^{12}$CO}

\def\arcsec{\ifmmode {^{\prime\prime}}\else $^{\prime\prime}$\fi}
\def\arcmin{\ifmmode {^{\prime}}\else $^{\prime}$\fi}

\def\simlt{\lower.5ex\hbox{\ltsima}}
\def\simgt{\lower.5ex\hbox{\gtsima}}
\def\ltsima{$\; \buildrel < \over \sim \;$}
\def\gtsima{$\; \buildrel > \over \sim \;$}
\def\ne2{[Ne\,II]}
\newcommand{\aap}{\rm A\&A}            
\newcommand{\apj}{\rm ApJ}                     
\newcommand{\apjl}{\rm ApJ}                
\newcommand{\aj}{\rm AJ}                       
\newcommand{\nat}{\rm Nature}                     
\newcommand{\pasp}{\rm PASP}                   
\newcommand{\mnras}{\rm MNRAS}                 

\begin{document}

\title{Understanding Mass-Loss and the Late Evolution of Intermediate Mass
Stars: Jets, Disks, Binarity, Dust, and Magnetic Fields}
\maketitle
\vspace{1in}
Dr. Raghvendra Sahai\\
Jet Propulsion Laboratory, 4800 Oak Grove Drive, Pasadena, CA\\
{\it email} sahai@jpl.nasa.gov, {\it phone} 818-354-0452\\
\vskip 0.2cm
\noindent Prof. Bruce Balick, University of Washington, Seattle, WA\\
Prof. Eric Blackman, University of Rochester, Rochester, New York, NY\\
Prof. Joel Kastner, Rochester Institute of Technology, Rochester, NY\\
Dr. Mark Claussen, National Radio Astronomy Observatory, AOC, Socorro, NM\\
Prof. Mark Morris, University of California at Los Angeles, Los Angeles, CA\\
Dr. Orsola De Marco, American Museum of Natural History, New York, NY\\
Prof. Angela Speck, University of Missouri, Columbia, MO\\
Prof. Adam Frank, University of Rochester, Rochester NY\\
Dr. Neal Turner, Jet Propulsion Laboratory, 4800 Oak Grove Drive, Pasadena, CA\\
\clearpage

\noindent {\bf 1. Background}\\
Almost all stars in the 1--8\,\ms~range evolve through the Asymptotic Giant Branch (AGB),
preplanetary nebula (PPN) and planetary nebula (PN) evolutionary phases. Most stars
that leave the main sequence in a Hubble time will end their lives in this way. The heavy mass loss
which occurs during the AGB phase is important across astrophysics, dramatically changing the
course of stellar evolution, dominantly contributing to the dust content of the interstellar medium
(ISM), and influencing its chemical composition (e.g., key biogenic elements like
$C$ and $N$). The particulate matter crucial for the birth of new solar systems is made
and ejected by AGB stars. The luminosity function of planetary nebulae is an important 
standard candle in determining cosmological distances (e.g., Jacoby et
al. 1996). Yet stellar evolution from the beginning of the AGB phase to the PN phase remains poorly
understood. First, we do not understand how the mass-loss (rate, geometry, temporal history)
depends on fundamental stellar parameters (stellar mass, luminosity, L; effective temperature,
$T_{eff}$, and
metallicity). Second, an entirely new dimension of complexity is added to this issue if one takes
into consideration the presence of a binary companion (common amongst pre-main-sequence
and main-sequence stars).

While the study of evolved non-massive stars has maintained a relatively modest
profile in recent decades, we are nonetheless in the midst of a quiet but exciting revolution in
this area, driven by new observational results, such as the discovery of jets and disks in stellar
environments where these were never expected and arcuate structures in outflowing circumstellar
envelopes produced on unanticipated time scales, and by the recognition of new symmetries such as
multipolarity and point-symmetry occuring frequently in the nebulae resulting from the outflows 
(e.g., see {\it conf. procs.} APN I--IV, 1994--2007). In
this paper we summarise the major unsolved problems in this field, and specify the areas where
allocation of effort and resources is most likely to help make significant progress.


\noindent {\bf 2. Major Unsolved Problems}\\
\noindent {\bf 2.1.} \underline{The AGB to post-AGB transition}\\
\noindent (i) {\it Aspherical Morphology \& Jets}: 
Many modern imaging surveys have shown that even though PPNs \& PNs evolve from the
spherically--symmetric mass-loss envelopes around AGB stars, the vast majority (\gsim\,80\%) of the
former deviate strongly from spherical symmetry  (e.g.,\,{Fig.\,\ref{starfish}). In an unbiased
survey of young PNs morphologies
with the Hubble Space Telescope (HST), Sahai \& Trauger (1998, ST98) found no round objects, but a
variety of bipolar and multipolar morphologies.
The significant changes in the circumstellar envelope (CSE) morphology during the evolutionary
transition
from the AGB to the post-AGB (pAGB) phase require a primary physical agent(s) which can break the
spherical symmetry of the radiatively-driven, dusty mass-loss phase. In the ``generalised
interacting-stellar-winds" (GISW) model, a fast ($>1000~km~s^{-1}$) isotropic wind from the PN
central star expands within an equatorially-dense AGB CSE (Balick 1987), and hydrodynamic
simulations (e.g., review by Balick \& Frank 2002) reproduce a variety of axisymmetric shapes. But
faced with the complexity,
organization and frequent presence of point-symmetry in the morphologies of their survey PNs, ST98
proposed that the primary agent for breaking spherical-symmetry is a jet or high-speed collimated
outflow (CFW) operating during the early post-AGB or late AGB evolutionary phase. The CFWs are
likely to be episodic, and either change their directionality (i.e., wobbling of axis or
precession) or have multiple components operating in different directions (quasi)simultaneously
(Sahai 2004). In
the ST98 model, primary shaping begins {\it prior} to the PN phase, and the variety of PN shapes
and structure depends in detail on the CFW characteristics (direction, strength, opening angle,
temporal history). If point-symmetric shapes result from the flow collimator precessing or becoming 
unstable, then what causes the destabilization?

Direct evidence for CFWs during the pre-PN phase has come from sensitive molecular line observations
which reveal the presence of very fast (few$\times$100 \kms) molecular outflows in PPNs and a few
very late AGB stars, with huge momentum-excesses which showed that these winds are not radiatively
driven (e.g. Bujarrabal et al. 2001, Sahai et al. 2006). Using STIS/HST, a carbon star, V
Hya, has been ``caught in the act" of ejecting a very fast (250\,\kms), highly collimated blobby
outflow (Sahai et al. 2003a). Strong support for the ST98 model was recently provided by a
(morphologically) unbiased HST imaging survey of young PPNs nebulae which shows very close
similarities in morphology between these objects and young PNs (Sahai et al. 2007). If the ST98
model is correct, then the question arises: what is the engine for producing CFW's? Can CFW's be
produced by single stars or is a binary companion essential? Single-star models have invoked
stellar rotation, strong magnetic fields, or both (e.g. Garcia-Segura et al 1999, Blackman et al.
2001), and binary models have invoked the angular momentum and/or the gravitational
influence of a companion (e.g. Morris 1987, Soker \& Livio 1994, Livio \& Pringle 1997). Yet,
in spite of vigorous debate (e.g., Bujarrabal et al. 2000), no consensus has yet emerged even as to
which of the above two broad classes of models is correct (Balick \& Frank 2002)! 

\noindent (ii) {\it Shock-Heated Bubbles}: 
X-ray (0.1-10 keV) observations have been used to probe the {\it direct signature of the fast and
slow wind interaction in PPNs and PNs}, i.e., the very hot ($\sim10^{7-8}$\,K) gas which should
be produced in the shocked region. Data from CXO and XMM have resulted in detections of nebular
emission in several PNs (e.g., Kastner et al. 2008 and refs therein) and one PPN
(He\,3-1475: Sahai et al. 2003b). These studies clearly show the presence of extended bubbles of hot
gas, in qualitative agreement with our expectations from interacting-winds models, but there is an
order of magnitude or more discrepancy between the expected and observed ($\sim$10$^6$\,K)
temperatures; and the presence of inhomogeneities in the X-ray surface brightness (BD+303639:
{Fig.\,\ref{starfish}) remain unresolved
(Soker \& Kastner 2003). Although conduction cooling might explain some cases (Steffen et al.
2008), it fails for the brightest and best-studied case, BD+303639 (Yu et al. 2009) for which
high-resolution X-ray spectra reveal the elemental abundances of the shock-superheated gas and
allow us to associate it with the undiluted, nucleosynthesis-enriched (C-rich) stellar wind. But
such a study has been successfully done in only this one object due to the
limited effective areas of CXO and XMM (300 ks yielded few$\times$100 cts in the brightest
lines). 

\noindent (iii) {\it Equatorial Waists \& Disks}: 
Most bipolar or multipolar PPNs or PNs harbor overdense, dusty equatorial waists. In PPNs, the
waists display convex (relative to the nebular center) sharp outer (radial) edges in absorption,
with radii typically $\gsim 1000$\,AU. Very infrequently, the waist appears as a disk with a sharp
outer boundary seen in emission at $few\times$100\,AU (e.g., IRAS04296 \& IRAS17106: Sahai 1999,
Kwok et al. 2000). The waist region can show complex symmetries (e.g, two torii in He2-113 \&
IRAS19024: Sahai et al. 2005a; point-symmetric microstructure in He2-47: Sahai 2000; central star
offset by $few\times$100\,AU from the symmetry-center in MyCn18: Sahai et al. 1999).

Significantly smaller equatorially-flattened structures or disks ($\sim$50\,AU) are found in a large
sub-class of young pAGB stars (de Ruyter et al. 2005), many of
which are RV Tauri stars (Jura 1986), and show photospheric depletion patterns similar to that seen
in the ISM (e.g., Maas et al. 2005). The depletion is
believed to result from a poorly understood process in which the circumstellar dust is trapped in a
disk, and the dust-depleted gas is accreted back onto the star (Waters et al. 1992). Mid-IR
spectroscopy with ISO and Spitzer has shown the ubiquitous presence of abundant cystalline
silicates in these objects implying the presence of
dust processing in long-lived disks with composition very similar to that seen in
planet-forming disks around young stars and in comet Hale-Bopp (Molster et al. 1999, Gielen et al.
2008). It has been
speculated that such disks may be sites where planets could form (Jura et al. 2006), as must
have occurred around pulsar PSR 1215+12 (Wolszczan \& Frail 1992).
But direct support for bound disks is limited
to one pAGB object: the Red Rectangle (closest known PPN with a binary central
star), where Bujarrabal et al. (2005), using interferometric CO(2-1) mapping, find a disk of outer
radius $\sim300-550$ AU. The disk mass (0.006\ms), dynamics, and temperature are amazingly similar
to protostellar disks in Herbig Ae/Be stars, in spite of its very different formation history.

The origin of the small, likely Keplerian, disks and the larger, dusty waists in pAGB objects is a
mystery, and the connection between these two (if any) is unknown. Compression of the AGB wind
towards the equatorial plane by a fast wind with a relatively large opening angle (Soker and
Rappaport 2000) is unlikely to work in most bipolar nebulae which appear to be highly-collimated,
momentum-driven shells. Although compact ($\lsim$1\,AU) disks form readily around a companion via
Bondi-Hoyle accretion of the primary AGB star's dusty wind (Mastrodemos and Morris 1998),
these simulations have so far not been able to produce the much larger circumbinary disks or the
dusty waists which we observe in pAGB objects.

Finally, a new puzzle has arisen with the discovery of large submm excesses in both the
disk-prominent pAGB objects (de Ruyter et al. 2005) as well as a few PPNs (Sahai et al. 2006, 
S{\'a}nchez Contreras et al. 2007), which
imply the presence of fairly substantial masses of very large ($\sim few\times100-1000$\,micron) 
grains. This discovery also opens up a new opportunity: namely the study of important physical
processes related to dust grain evolution such as coagulation in an environment similar to, but
much simpler than protostellar disks. PAGB disks will allow us to probe the very early stages of
grain coagulation, i.e. on time-scales (1000 years), not possible with studies of   planet-forming
disks which are typically $\gsim10^6$\,yr old.
 
\noindent {\bf 2.2.} \underline{Binarity: Mass-Loss \& Evolution}:
Binarity can strongly influence mass-loss in evolved stars. Perhaps all asymmetric PNs and PPNs
involve binaries (Moe \& de Marco 2006; Soker 2006), but testing such a hypothesis requires a far
better knowledge of the incidence of binarity in these objects than available currently. Direct
observational evidence of binarity has been very hard to come by due to observational limitations.
AGB stars are very luminous ($\sim$\,10$^3$-10$^4$\,\ls), and surrounded by dusty envelopes, making
it very difficult to directly detect the light from nearby stellar companions which are generally
likely to be significantly less luminous main-sequence stars or white dwarfs. Radial-velocity or
photometric variability measurements are not feasible for AGB stars due to their intrinsic large
variability and pulsations.
In the case PN central stars, photometric variability measurements (Bond 2000) have met with limited
success. Extensive observations over decades have led to a sum total of $\lsim$20 binary CSPN (Bond
2000, Ciardullo et al. 1999), implying a 10-15\% fraction of detectable close binaries among
randomly selected PNs. New efforts by de Marco (2009) have found variable radial velocities
in 10/11 PN central stars, but these lack definitive confirmation as binaries due to stellar
variability. {\it Bond concludes that it is likely that the known short-period binaries in PNs are
only the tip of an iceberg of a substantial population of longer-period binaries}, given the
limitations of the search techniques which reveal only binaries with very short periods
($\lsim$\,few days). In the case of PPNs, radial-velocity measurements (Hrivnak et al. 2008) have
not been succesful due to pulsational stellar variability.

Binarity may cut short the primary's AGB
evolution as appears to be the case for disk-prominent pAGB stars, where the primary and
companion stars are presently not in contact, but
show orbits which are too small to accommodate a full-grown AGB star, and the disk sizes are large
enough that they must be circumbinary. But no theoretical models exist
to explain the formation of these systems.

\noindent {\bf 2.3.} \underline{AGB Mass-Loss}: The dusty, spherical, radiatively-driven winds from
cool
AGB stars are now well-observed via molecular line emission (e.g., review by Olofsson 2006). The
CSE dust component can be seen out to very large radii in scattered light -- HST images of
the Egg Nebula and other evolved stars trace the dusty AGB wind out to $few\times10^4$ yrs.
One of the dramatic, but poorly understood phenomena uncovered by such high-resolution images is
the presence of roughly concentric arcs in the AGB envelopes of a small number of objects (e.g.,
review by Su 2004). Theory is catching up with observations -- numerical simulations of dust
condensation in C-rich ($C/O\,>\,1$) AGB atmospheres produce dust-driven winds broadly consistent
with observations (e.g., Nowotny et al. 2005), but the same is not true of models of silicate dust
in O-rich
stars (e.g.,  H{\"o}fner 2008). Crucial input for such simulations has started to come from near-
and
mid-IR interferometry (e.g., with ISI, VLTI/\,MIDI \& AMBER, and the Keck Interferometer: e.g.
Danchi et al. 1990, Ohnaka et al. 2008, Mennesson et al. 2005), which have provided us with glimpses
into the dust-condensation regions where the dusty outflow is accelerated.
But a prescription for mass loss rates and gas-to-dust ratios as a function of fundamental stellar
parameters and evolutionary phase remains a distant goal. One of the hurdles in making progress on
this front has been the lack of effort in inferring accurate gas and dust mass loss rates from the
data (Sahai 1990). Determining the dust-to-gas ratio in a variety of objects (C-rich, O-rich and
S-type) is crucial to the determination of total mass-loss rates because observationally one can
detect mass-loss via thermal dust emission to much larger distances and much lower rates than
thermal
molecular line emission.  

\noindent {\bf 3. Future Research}\\
We describe below a sampling of observational, theoretical and computational studies which are
likely to make the maximum impact in the coming decade on the problems described above.\\ 
\noindent {\bf 3.1.} \underline{Observational}\\
\noindent$\bullet${\it Optical/NIR}
Near-IR and mid-IR interferometric instruments, specially those which provide closure-phase data
(VLTI/\,AMBER, ISI/Mt.Wilson), low-resolution (R$\sim$530) spectral data (VLTI/\,MIDI), and real
images (upcoming 6-element reconfigurable Magdalena Ridge Observatory Interferometer [MROI], New
Mexico, first light in 2010), with resolutions ranging from 0.1 to 100, can be used to probe the
launch regions of CFWs in late AGB stars and PPNs, in particular the disk temperature, geometry and
density structure. Direct imaging, with space-based telescopes such as HST and
JWST will remain unsurpassed in providing large field-of-view images with very high dynamical range
(crucial for detecting faint circumstellar structures next to bright central stars) at $\sim$100
mas resolution, because of their very stable PSFs. Multi-epoch high-resolution images spaced by
5-10\,yr can be used to trace nebular proper motions (e.g., Sahai et al. 2002).
The ability to probe kinematics at high spatial
resolution is crucial to probing jet acceleration and kinematics close to the launch site, and can
be carried out from the ground in the NIR with integral-field (e.g., OSIRIS/Keck) or long-slit
spectrographs  (e.g., NIRSPEC/Keck) behind AO; and in the optical if STIS/HST is repaired. The
study of astrophysical disks and jets, not only in PPNs and PNs, but across most astrophysical
environments, can make great strides if a high-resolution ($\sim$10\,\kms) long-slit or
integral-field spectrograph becomes available on a future space-based optical/UV platform. 

\noindent$\bullet${\it Radio and Millimeter-Wave}
Radio and mm interferometry provide an unsurpassed combination of high angular resolution and high
velocity resolution for studies of AGB stars, PPNs and PNs. Key objects for such studies are 
the ``water-fountain" PPNs, which are distinguished by the presence of very high-velocity red- and
blue-shifted H$_2$O and/or OH maser features. These are arguably amongst the youngest PPNs and
therefore most likely to show active jet sculpting (e.g., 
IRAS16342 shows H$_2$O maser features with radial velocities separated by more than 250\,\kms
and the telltale ``corkscrew" signature of a precessing jet in near-IR
AO/Keck images: Sahai et al. 2005b). The H$_2$O masers in these sources lie on the opposite sides of
a bipolar jet, and being very bright and compact, are unsurpassed tracers of the jet proper motions 
(e.g., Claussen et al. 2009). Such proper motion (and radial velocity) measurements made
with the NRAO Very Long Baseline Array (VLBA) over several epochs can a) determine the geometric
parallax of sources with unparalleled accuracy to 3 kpc, and beyond (c.f. Imai et al. 2007) and b)
determine the high-velocity jet's 3-D motions (e.g., whether or not the jet is precessing --
Imai et al. 2002). High spectral resolution monitoring using the Green Bank Telescope can
accurately determine accelerations and decelerations in the jet radial velocity. Such measurements
will greatly constrain theoretical models of how these high velocity jets sculpt the circumstellar
medium in the evolution of these objects to PNs. Upgrades to the sensitivity of
the VLBA will enhance the continuum sensitivity to extragalactic source as fiducial markers for the
the measurement of proper motions, and allow geometric parallaxes to be measured to 10 or 15 kpc
over a 1-year period. This improvement will not only facilitate parallax measurements for
H$_2$O-fountain PPNs, but for optically invisible dust-obscured AGB stars as well, which harbor
masers in their shells.

Thermal molecular gas emission in the outflows of these transition objects
will be imaged by the ALMA in the coming decade. At 100 mas resolution in mm-wave molecular
transitions of high-density tracers, we should see evidence for the high-velocity jet's interaction 
with the remnant AGB gas at $\sim$100 AU (or better, depending upon the distance)
from the star. ALMA will be very important for finding and studying Keplerian disks in
disk-prominent pAGB objects, and together with the EVLA, radio/mm/submm continuum emission from
the large grain component in these disks. ALMA should also be able to probe magnetic fields in the
disks and outflow in PNs and PPNs via the presence of submm polarisation resulting from aligned
grains (e.g., detected via large-beam observations in a few 
objects: Sabin et al. 2007).


\noindent$\bullet${\it X-Ray} Surveys of 
large PN and PPN samples with increased sensitivity (e.g., with the International X-ray
Observatory) can produce enormous scientific impact in areas as wide-ranging as
the structural evolution of PNs, the physics of wind shocks, and our understanding of galactic
chemical evolution. Detections of X-rays towards the central stars in PPNs can provide us
with a unique probe of magnetic fields in the outflow engine vicinity 
(dissipation of a solar analog dynamo-generated field should result in a non-thermal X-ray
luminosity of 10$^{32}$ erg\,s$^{-1}$ from the central stars in very young PPNs [Blackman et al.
2001]).

\noindent$\bullet${\it Binary Searches \& Astrometry}
Ultraviolet photometry (e.g., with GALEX) is a promising new technique for 
finding hot ($T_{eff}\gsim$7000\,K) companions of cool AGB stars in large numbers (Sahai et al.
2008). 
Astrometric missions such as GAIA mission will result in the discovery of large numbers of
binary AGB stars, and possibly orbital characterisation of some fraction of these. In conjunction
with GAIA, LSST will be able to robustly establish a period-luminosity relationship
for Galactic long period variable stars ascending the AGB. Although 
GAIA will provide distances to a very large number of dying stars, it is a
survey mission and does not allow one to design the observational cadence necessary for probing
binaries with the relevant spread of periods. Further, high mass-loss
rate objects (\gsim\,5$\times\,10^{-6}$\,\my), which dominate the recycling of processed matter
to the ISM (Jura \& Kleinmann 1989) are heavily enshrouded by
dust and are rather faint (m$_V$\,\gsim\,15). GAIA's sensitivity (25 $\mu$sec for
objects with $V\lsim15$ mag) will be adversely affected by the presence of extended faint
nebulosity. The Space Interferometry Mission (SIM) is much less sensitive to this
effect, because its interferometric technique results in loss of sensitivity to structures that
are smooth on scales more extended than about 10 mas. Thus, for the search for binarity in
optically-faint, heavy mass-loss AGB stars, as well as the central stars of PPNs and PNs, the
resurrection of SIM is crucial.

\noindent {\bf 3.2.} \underline{Theory and Modelling}:\\
\noindent$\bullet${\it Central Engines \& Fast Collimated Outflows}  
The next decade of theoretical research must go beyond kinematics to determine specific testable
predictions for different MHD jet-engine scenarios, including the relative importance of isolated
versus binary central engines. The large momentum excesses in PPNs outflows is an important clue to
figuring out the physical mechanism powering the outflows, and likely ones should involve
those that draw power from the engine's rotational energy. Just radiation and rotation are
insufficient but scenarios
involving a symbiosis between rotation, differential rotation, accretion, and dynamo-produced large
scale magnetic fields are promising (Blackman et al. 2001, Matt et al. 2006). Differential rotation
supplied by
binaries can amplify magnetic fields that can in turn produce accretion-powered bipolar jets.
Binary mechanisms that power PPNs and PNs include common envelope scenarios that can produce a
combination of asymmetric features (Nordhaus \& Blackman 2006). Direct, equatorially-enhanced
hydrodynamic mass ejection (e.g., due to rotation) can produce the dusty waists in PPNs, whereas
envelope dynamos + MHD outflows or dynamos in accretion disks can in principle produce strong
poloidal outflows. The unshocked CFW in PPNs and PNs (e.g., as in He\,2-90: Sahai et al. 2002) is 
rarely seen directly, and hydro simulations (accounting for low-temperature cooling and ionisation
effects) fitted to observations are needed to infer their properties (e.g., Lee \& Sahai 2003,2004,
Akashi \& Soker 2008).

\noindent$\bullet${\it Making Disks} 
Accretion disks may form from a shredded low mass companion around an AGB primary core (Reyes-Ruiz
\& Lopez, 1999), or around the secondary (Morris 1987, Soker and Livio 2004). Huggins (2007) finds
that dense waists in PPNs and PNs develop a few hundred years before jets, supporting the
interacting binary scenario for producing a bound disk. Smoothed-particle hydrodynamics (SPH) codes
developed to investigate gas flows in binaries with a mass-losing primary 
(Mastrodemos and Morris 1998, 1999) have demonstrated how unbound outflow from
the primary can be concentrated toward the system's equatorial plane into a small disk (and how
the density structure in the extended outflow forms an archimedean spiral shell, which can explain 
the remarkable circumstellar spiral structure seen in the CSE of the extreme carbon star,
CRL\,3068: Morris et al. 2006, Mauron \& Huggins 2006). New
simulations with such codes covering a larger input parameter space, and incorporating improved
physics need to be carried out in order to see if bound, circumbinary disks can be formed, and if
these turn out to be very extended in specific cases, then the sharp disk boundaries directly
seen in some PPNs could be explained (e.g., due to a very distant giant planet or brown dwarf
companion).

\noindent$\bullet${\it AGB Mass-Loss Rates and Gas-to-Dust Ratios} For an accurate determination of
mass-loss rates and dust-to-gas ratios, a self-consistent treatment of thermodynamics, radiative
transfer and dust emission (e.g., as in Sahai 1990) needs to be applied to the molecular-line and
dust emission observations of a large number of mass-losing stars, covering the basic abundance
types (C-rich, O-rich and S-type) and going down to the lowest mass-loss rate envelopes observable
in both gas and dust emission. The thermodynamical computation requires the development of fast
methods for computing the radiative cooling due to the major molecular coolants (CO and H$_2$O).
For the highly aspherical molecular outflows in PPNs, 2-D and possibly 3-D line radiative transfer
codes, including heating-cooling, needs to be developed. So far simple, only 2D models with a
constant temperature and LTE conditions have been attempted.

\vskip -0.15in
\begin{figure}[htbp]
\vskip 0.1in
\resizebox{0.28\textwidth}{!}{\includegraphics{r_j_h_pap.epsi}}
\resizebox{0.28\textwidth}{!}{\includegraphics{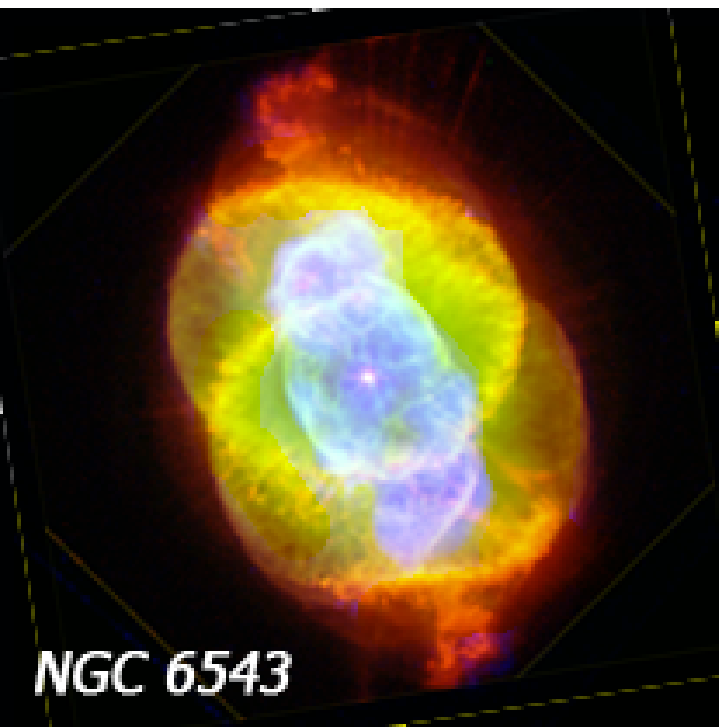}}
\resizebox{0.28\textwidth}{!}{\includegraphics{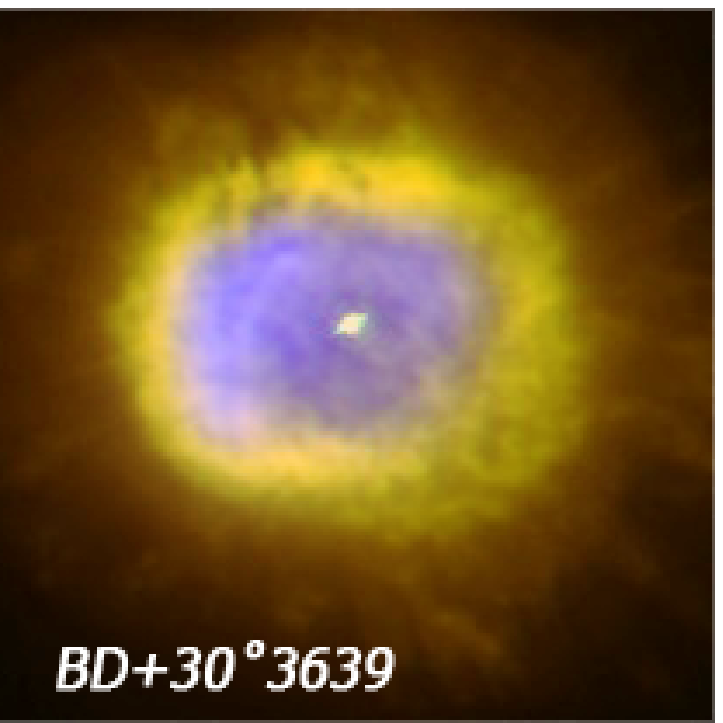}}
\vskip -0.1in
\caption{({\it left}) The multipolar PPN IRAS19024 (scattered light HST image in R, H \& K-band) 
and PNs ({\it cen.}) NGC6543 \& ({\it right}) BD+30\arcdeg3639 (optical
forbidden-line emission [HST] and X-ray emission [CXO]).
}
\label{starfish}
\end{figure}
\vskip -0.1in
In summary, the next decade, especially with the advent of JWST, ground-based interferometers such
as ALMA, EVLA, MROI, and survey telescopes (LSST), holds great promise for  
patching some very large holes in the theory of stellar evolution and mass-loss on and beyond the
AGB, testing the physics of launching MHD stellar jets and outflows and the formation of dusty
disks under ideal conditions, and the role of binarity in these important problems.

\footnotesize
\begin{minipage}{8.5cm}
\noindent {\bf References}:\\
$\bullet$ Akashi, M., \& Soker, N.\ 2008, \mnras, 391, 106\\
$\bullet$ Asymmetric Planetary Nebulae (APN) I, 1995, eds. A. Harpaz \& N. Soker, Ann. Israel Phys.
Soc., 11\\
$\bullet$ Asymmetric Planetary Nebulae (APN) II, 2000, eds. J.H. Kastner, N. Soker, \& S. Rappaport,
ASP Conf. Ser. 199\\
$\bullet$ Asymmetric Planetary Nebulae (APN) III, 2004, eds. M. Meixner et al., ASP Conf. Ser. 313\\
$\bullet$ Asymmetric Planetary Nebulae (APN) IV, 2007, eds. R.L.M. Corradi, A. Manchado, N. Soker\\
$\bullet$ Balick, B.\ 1987, \aj, 94, 671\\ 
$\bullet$ Balick, B.~\& Frank, A. 2002, ARAA, 40, 439\\
$\bullet$ Blackman, E.~G., 
Frank, A., \& Welch, C.\ 2001, \apj, 546, 288\\ 
$\bullet$ Bond, H.~E.\ 2000, APN II, 199, 115\\
$\bullet$ Bujarrabal, V. et al. 2000, APN II, 199, 201\\
$\bullet$ Bujarrabal, V. et al. 2001, A\&A, 377, 868\\
$\bullet$ Bujarrabal, V. et al. 2003, A\&A, 409, 573\\ 
$\bullet$ Bujarrabal, V. et al. 2005, \aap, 441, 1031\\
$\bullet$ Claussen, M.~J., Sahai, R., \& Morris, M.~R.\ 2009, \apj, 691, 219\\ 
$\bullet$ Danchi, W.~C. et al. 1990, \apjl, 359, L59\\
$\bullet$ de Marco, O. 2009, arXiv:0902.1137\\
$\bullet$ de Ruyter, S., van Winckel, H., Dominik, C., et al. 2005, A\&A, 435, 161\\
$\bullet$ Garc{\'{\i}}a-Segura, G. et al. 1999, \apj, 517, 767\\
$\bullet$ Gielen, C. et al. 2008, \aap, 490, 725\\
$\bullet$ H{\"o}fner, S.\ 2008, \aap, 491, L1\\
$\bullet$ Huggins, P.~J.\ 2007, \apj, 
663, 342\\ 
$\bullet$ Imai, H. et al. 2002, Nature, 417, 829\\
$\bullet$ Imai, H., Sahai, R. \& Morris, M. 2007, ApJ, 668, 424\\
$\bullet$ Jacoby, G.~H., Ciardullo, R., \& Harris, W.~E.\ 1996, \apj, 462, 1\\
$\bullet$ Jura, M. 1986, ApJ, 309, 732\\
$\bullet$ Jura, M., \& Kleinmann, S.~G.\ 1989, \apj, 341, 359\\
$\bullet$ Jura, M., et al.\ 2006, \apjl, 637, L45\\
$\bullet$ Kastner, J.H. et al. 2008, \apj, 672, 957\\
$\bullet$ Kwok, S., Hrivnak, B.~J., \& Su, K.~Y.~L.\ 2000, \apjl, 544, L149\\
$\bullet$ Likkel, Morris \& Maddalena 1992, A\&A, 256, 581\\
$\bullet$ Livio, M., \& Pringle, J.~E.\ 1997, \apj, 486, 835\\
$\bullet$ Maas, T., Van Winckel, H., \& Lloyd Evans, T.\ 2005, \aap, 429, 297\\
$\bullet$ Matt, S., Frank, A., \& Blackman, E.~G.\ 2006, \apjl, 647, L45\\
$\bullet$ Mauron, N., \& Huggins, P.~J.\ 2006, \aap, 452, 257\\
$\bullet$ Mastrodemos, N., \& Morris, M.\ 1998, \apj, 497, 303\\
$\bullet$ Mastrodemos, N., \& Morris, M.\ 1999, \apj, 523, 357\\
\end{minipage}

\vspace{-20cm}
\hspace{8.8cm}
\begin{minipage}{8.5cm} \vspace{-0.5cm}
$\bullet$ Mennesson, B., et al.\ 2005, \apjl, 634, L169\\
$\bullet$ Moe,  M. \&   De Marco, O.,  2006. ApJ, 650, 916\\
$\bullet$ Morris, M.\ 1987, \pasp, 99, 1115\\ 
$\bullet$ Morris, M. et al. 2006, IAU Symp. 234, 469\\
$\bullet$ Nordhaus, J., \& Blackman, E.G., 2006, MNRAS, 370, 2004\\
$\bullet$ Nowotny, W., Lebzelter, T., Hron, J., H{\"o}fner, S.\ 2005, \aap, 437, 285\\
$\bullet$ Ohnaka, K., Izumiura, H., Leinert, C.et al. 2008, \aap, 490, 173\\ 
$\bullet$ Olofsson, H.\ 2006, Reviews in Modern Astronomy 19 (Wiley-VCH Verlag, Wenheim), pp. 75\\ 
$\bullet$ Pollack, J.~B., Hollenbach, D., Beckwith, S., et al. 1994, ApJ, 421, 615\\
$\bullet$ Reyes-Ruiz, M., \& Lopez, J. A. 1999 ApJ, 524, 952\\
$\bullet$ Sahai, R.\ 2000, \apjl, 537, L43\\
$\bullet$ Sahai, R. 2004, ASP Conf.\,Ser., 313, 141\\ 
$\bullet$ Sahai, R. \& Trauger, J.T. 1998, AJ, 116, 1357\\
$\bullet$ Sahai, R. et al.\ 1999, \aj, 118, 468\\
$\bullet$ Sahai, R. et al. 2002, ApJ, 573, L123\\
$\bullet$ Sahai, R. et al. 2003a, \nat,  426, 261\\
$\bullet$ Sahai, R. et al. 2003b, \apjl, 599, L87\\
$\bullet$ Sahai, R., S{\'a}nchez Contreras, C., \& Morris, M.\ 2005a, \apj, 620, 948\\
$\bullet$ Sahai, R. et al. 2005b, \apjl, 622, L53\\
$\bullet$ Sahai, R., Young, K., Patel, N., et al. 2006, \apj, 653, 1241\\ 
$\bullet$ Sahai, R. et al. 2007, \aj, 134, 2200\\
$\bullet$ Sahai, R., Findeisen, K., de Paz, A.~G., \& S{\'a}nchez Contreras, C.\ 2008, \apj, 689,
1274\\
$\bullet$ Sabin, L., Zijlstra, A.~A., \& Greaves, J.~S.\ 2007, ASP, 378, 337\\ 
$\bullet$ S{\'a}nchez Contreras, C., et al. 2007, \apj, 656, 1150\\ 
$\bullet$ Soker, N. 2001, MNRAS, 328, 1081\\
$\bullet$ Soker, N. 2006, ApJ, 645, L57\\
$\bullet$ Soker, N., \& Kastner, J.~H.\ 2003, \apj, 583, 368\\
$\bullet$ Soker, N., \& Livio, M.\ 1994, \apj, 421, 219\\
$\bullet$ Soker, N., \& Livio, M., 2004,  ApJ, 421, 219\\
$\bullet$ Soker, N., \& Rappaport, S.\ 2000, \apj, 538, 241\\
$\bullet$ Sopka, R.J., Hildebrand, R., Jaffe, D.T. et al. 1985, ApJ, 294, 242\\
$\bullet$ Steffen, M. et al. 2008, A\&A, 489, 173\\
$\bullet$ Su, K.Y.L. 2004, APN III, 313, 247\\
$\bullet$ Van Winckel, H. 2003, ARA\&A 41, 391\\
$\bullet$ Waters, L.B.F.M., Trams, N.R., \& Waelkens, C.\ 1992, A\&A, 262, L37\\
$\bullet$ Wolszczan, A., \& Frail, D.~A.\ 1992, \nat, 355, 145\\
$\bullet$ Yu, Y. et al. 2009, ApJ, 690, 440\\
\end{minipage}

\end{document}